# Evaluating the Dynamic Behavior of Database Applications

Zhen He[1] and Jérôme Darmont[2]

[1] Department of Computer Science

University of Vermont

Burlington, VT 05405

USA

Tel: +1 802 656 3815

Fax: +1 802 656 0696

zhenhe@emba.uvm.edu

[2] ERIC, Université Lumière Lyon 2

5 avenue Pierre Mendès-France

69676 Bron Cedex

France

Tel: +33 478 774 403

Fax: +33 478 772 375

jerome.darmont@univ-lyon2.fr

# Evaluating the Dynamic Behavior of Database Applications

**Abstract**

This paper explores the effect that changing access patterns has on the performance of database management systems. Changes in access patterns play an important role in determining the efficiency of key performance optimization techniques, such as dynamic clustering, prefetching, and buffer replacement. However, all existing benchmarks or evaluation frameworks produce static access patterns in which objects are always accessed in the same order repeatedly. Hence, we have proposed the Dynamic Evaluation Framework (DEF) that simulates access pattern changes using configurable styles of change. DEF has been designed to be open and fully extensible (e.g., new access pattern change models can be added easily). In this paper, we instantiate DEF into the Dynamic object Evaluation Framework (DoEF) which is designed for object databases, i.e., object-oriented or object-relational databases such as multi-media databases or most XML databases.

The capabilities of DoEF have been evaluated by simulating the execution of four different dynamic clustering algorithms. The results confirm our analysis that flexible conservative re-clustering is the key in determining a clustering algorithm's ability to adapt to changes in access pattern. These results show the effectiveness of DoEF at determining the adaptability of each dynamic clustering algorithm to changes in access pattern in a simulation environment. In a second set of experiments, we have used DoEF to compare the performance of two real-life object stores : Platypus and SHORE. DoEF has helped to reveal the poor swapping performance of Platypus.

## Introduction

Performance evaluation is critical for both designers of Database Management Systems (DBMSs), for architectural or optimisation choices, and users, for efficiency comparison or tuning. Traditionally, this is achieved with the use of benchmarks, i.e., synthetic workload models (databases and operations) and sets of performance metrics. Although in real life, almost no application always accesses the same data in the same order repeatedly, none of the existing database benchmarks incorporate the possibility of change in the access patterns. The ability to adapt to changes in access patterns is critical to database performance. In addition, highly tuning a database to perform well for only one particular access pattern can lead to poor performance when different access patterns are used. Thus, a database tuned to a particular trace (a particular instance of a real application usage) is likely to perform poorly when a different trace is used. In addition, the performance of a database on a particular trace provides little insight into the reasons behind its performance, and thus is of limited use to database researchers or engineers, who are interested in the identification and improvement in the performance of particular components of the system.

Thus, the aim of our work is to provide a means for them to explore the performance of databases under different styles of access pattern change. In contrast, benchmarks of the TPC family aim to provide standardised means of comparing systems for vendors and customers. In this paper, we take a first look at how dynamic application behavior can be modeled and propose the Dynamic Evaluation Framework (DEF). DEF makes the first attempt at exploring the issue of evaluating the performance of DBMSs in general and such optimization techniques as dynamic clustering algorithms in particular, with respect to changing query profiles. DEF contains a set of protocols which in turn define a set of styles of access pattern change. DEF by no means has exhausted all possible styles of access pattern change. However, we have designed DEF to be fully extensible and its design allows new styles of change to be easily incorporated. Finally, DEF is a generic platform that can be specialized to suit the particular needs of a given family of DBMS (e.g., relational, object, or object-relational). In particular, it is designed to be implemented on top of an existing benchmark so that previous benchmarking research and standards can be reused.

In this paper, we show the utility of DEF by creating an instance of DEF called the Dynamic object Evaluation Framework (DoEF) (He and Darmont, 2003) which is designed for object databases.

Note that, in the remainder of this paper, we term Object Database Management Systems (ODBMSs) both object-oriented and object-relational systems, indifferently. ODBMSs include most multimedia and XML DBMSs, for example. DoEF is built on top of the Object Clustering Benchmark (OCB) (Darmont, Petit, and Schneider, 1998; Darmont and Schneider, 2000), which is a generic object-oriented benchmark that is able to simulate the behavior of other main object-oriented benchmarks. DoEF uses both the database built from the rich schema of OCB and the operations offered by OCB. Since OCB's generic model can be implemented within an object-relational system and most of its operations are relevant for such a system, DoEF can also be used in the object-relational context.

To test the effectiveness of DoEF, we have conducted two sets of experiments. First we have benchmarked four state of the art dynamic clustering algorithms (Bullat and Schneider, 1996; Darmont, Fromantin, Regnier, Gruenwald, and Schneider, 2000; He, Marquez, and Blackburn, 2000). There are three reasons for choosing to test the effectiveness of DoEF using dynamic clustering algorithms: (1) ever since the "early days" of object database management systems, clustering has been proven to be one of the most effective performance enhancement techniques (Gerlhof, Kemper, and Moerkotte, 1996); (2) the performance of dynamic clustering algorithms is very sensitive to changing access patterns; and (3) despite this sensitivity, no previous attempt has been made to benchmark these algorithms in this way. Then we tested the utility of DoEF by benchmarking two transactional object stores: Platypus (He, Blackburn, Kirby, and Zigman, 2000); and SHORE (Carey, DeWitt, Franklin, Hall, McAuliffe, Naughton, Schuh, Solomon, Tan, Tsatalos, White, and Zwilling, 1994).

Our first paper about DoEF (He and Darmont, 2003) made two key contributions: (1) it proposed the first evaluation framework that allowed ODBMSs and associated optimisation techniques to be evaluated in a dynamic environment; (2) it presented the first performance evaluation experiments of dynamic clustering algorithms in a dynamic environment (by simulation). This paper expands on this material by presenting a more generic view of our evaluation framework, by providing a more thorough description of the configurable styles of change, and by reporting the results of new experiments that validate the effectiveness of DoEF at contrasting the dynamic performance of two real-life ODBMSs.

The remainder of this paper is organised as follows. We first present a brief description of existing DBMS benchmarks. Second we present an overview of the OCB benchmark. The next two sections describe in detail the DEF framework and its object-oriented instance DoEF, respectively. Next we presents a brief description of the state of the art clustering algorithms and object stores we have used in this paper. We present and discuss experimental results achieved with DoEF in the next section, and finally conclude the paper and provide future research directions.

## Existing Benchmarks

We briefly describe here the prevalent benchmarks, besides OCB that is detailed in the next section, which have been proposed in the literature for evaluating the performances of DBMSs. Note that none of these benchmarks incorporate any dynamic application behavior.

In the world of relational databases, the Transaction Processing Performance Council (TPC), a non-profit institute founded in 1988, defines standard benchmarks, verifies their correct application, and publishes the results. The TPC benchmarks include TPC-C (TPC, 2002a) for OLTP, TPC-H (TPC, 2003a) and TPC-R (TPC, 2003b) for decision support, and TPC-W (TPC, 2002b) for web commerce. All these benchmarks feature an elaborate database and set of operations. Both are fixed, the only parameter being the database size (scale factor).

In contrast, there is no standard object-oriented database benchmark. However, the OO1 benchmark (Cattell, 1991), HyperModel benchmark (Anderson, Berre, Mallison, Porter, and Scheider, 1990), and the OO7 benchmark (Carey, DeWitt, and Naughton, 1993) may be considered as de facto standards. They are all designed to mimic engineering applications such as CAD, CAM, or CASE applications. They range from OO1, that has a very simple schema (two classes) and only three simple operations, to OO7, that is more generic and provides both a much richer and more customisable schema (ten classes), and a wider range of operations (fifteen complex operations). However, even OO7's schema is static and still not generic enough to model other types of applications like financial, telecommunications and multimedia applications (Tiwary, Narasayya, and Levy, 1995). Furthermore, each step in adding complexity makes these benchmarks harder to implement.

Object-relational benchmarks, such as the BUCKY benchmark (Carey, DeWitt, Naughton, Asgarian, Brown, Gehrke, and Shah, 1997) and Benchmark for Object-Relational Databases (BORD) (Lee, Kim, and Kim, 2000), are query-oriented benchmarks that are specifically aimed at evaluating the performances of object-relational database systems. For instance, BUCKY only features operations that are specific to object-relational systems, since typical object navigation has already been tested by other benchmarks (see above). Hence, these benchmarks focus on queries involving object identifiers, inheritance, joins, class references, inter-object references, set-valued attributes, flattening queries, object methods, and various abstract data types. The database schema is also static in these benchmarks.

Finally, Carey and Franklin have designed a set of workloads for measuring the performance of their client-server Object-Oriented Database Management Systems (OODBMSs) (Carey, Franklin, Livny, and Shekita, 1991; Franklin, Carey, and Livny, 1993). These workloads operate at the page grain instead of the object grain, i.e., synthetic transactions read or write pages instead of objects. The workloads contain the notion of hot and cold regions (some areas of database are more frequently accessed compared to others), attempting to approximate real application behaviour. However, the hot region never moves, meaning no attempt is made to model dynamic application behaviour.

## The Object Clustering Benchmark (OCB)

OCB is a generic, tunable benchmark aimed at evaluating the performances of OODBMSs. It was first oriented toward testing clustering strategies (Darmont et al., 1998) and was later extended to become fully generic (Darmont and Schneider, 2000). The flexibility and scalability of OCB is achieved through an extensive set of parameters. OCB is able to simulate the behavior of the de facto standards in object-oriented benchmarking, namely OO1 (Cattell, 1991), HyperModel (Anderson et al., 1990), and OO7 (Carey et al., 1993). Furthermore, OCB's generic model can be implemented within an object-relational system easily and most of its operations are relevant for such a system. We only provide here an overview of OCB. Its complete specification is available in (Darmont and Schneider, 2000). The two main components of OCB are its database and workload.

**Database**

The OCB database is made up of NC classes derived from the same metaclass (Figure 1). Classes are defined by two parameters: MAXNREF, the maximum number of references in the instances and BASESIZE, an increment size used to compute the InstanceSize. Each CRef (class reference) has a type: TRef. There are NTREF different types of references (e.g., inheritance, aggregation...). Finally, an Iterator is maintained within each class to save references toward all its instances.

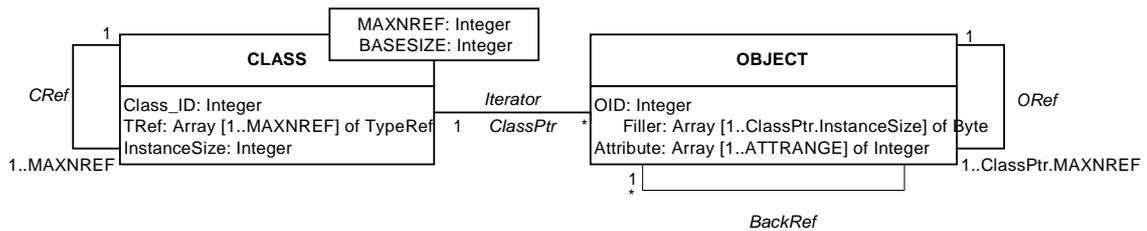

Fig. 1. OCB database schema

Each object possesses ATTRANGE integer attributes that may be read and updated by transactions. A Filler string of size InstanceSize is used to simulate the actual size of the object. After instantiating the schema, an object O of class C points through the ORef references to at most C.MAXNREF objects. There is also a backward reference (BackRef) from each referenced object toward the referring object O.

The database generation proceeds through three steps.

1. Instantiation of the CLASS metaclass into NC classes and selection of class level references. Class references are selected to belong to the [Class_ID - CLOCREF, Class_ID + CLOCREF] interval. This models locality of reference at the class level.

2. Database consistency check-up: suppression of all cycles and discrepancies within the graphs that do not allow them, e.g., inheritance graphs or composition hierarchies.
3. Instantiation of the NC classes into NO objects and random selection of the object references. Object references are selected to belong to the [OID - OLOCREF, OID + OLOCREF] interval. This models locality of reference at the instance level.

The main database parameters are summarized in Table 1.

| Parameter name | Parameter | Default value |
| --- | --- | --- |
| NC | Number of classes in the database | 50 |
| MAXNREF(i) | Maximum number of references, per class | 10 |
| BASESIZE(i) | Instances base size, per class | 50 bytes |
| NO | Total number of objects | 20,000 |
| NREFT | Number of reference types | 4 |
| ATTRANGE | Number of integer attributes in an object | 1 |
| CLOCREF | Class locality of reference | NC |
| OLOCREF | Object locality of reference | NO |

Table 1. OCB database main parameters

**Workload**

The operations of OCB are broken up into four categories.

– Random Access: Access to NRND randomly selected objects.
– Sequential Scan: Randomly select a class and then access all its instances. A Range Lookup additionally performs a test on the value of NTEST attributes, for each accessed instance.
– Traversal: There are two types of traversals in OCB. Set-oriented accesses (or associative accesses) perform a breadth-first search. Navigational Accesses are further divided into Simple Traversals (depth-first searches), Hierarchy Traversals that always follow the same reference type, and Stochastic Traversals that select the next link to cross at random. Each traversal proceeds from a randomly chosen root object, and up to a predefined depth. All the traversals can be reversed by following the backward links.

– Update: Update operations are also subdivided into different types. Schema Evolutions are random insertions and deletions of Class objects (one at a time). Database Evolutions are random insertions and deletions of objects. Attribute Updates randomly select NUPDT objects to update, or randomly select a class and update all of its objects (Sequential Update).

## The Dynamic Evaluation Framework (DEF)

The primary goal of DEF is to evaluate the dynamic performance of DBMSs. To make the work of DEF more general, we have made two key decisions: define DEF as an extensible framework; and reuse existing and standard benchmarks when available.

### Dynamic Framework

We start by giving an example scenario that the framework can mimic. Suppose we are modeling an on-line book store in which certain groups of books are popular at certain times. For example, travel guides to Australia during the 2000 Olympics may have been very popular. However, once the Olympics is over, these books may suddenly or gradually become less popular.

Once the desired book has been selected, information relating to the book may be required. Example required information includes customer reviews of the book, excerpts from the book, picture of the cover, etc. If the data are stored in an ODBMS, retrieving the related information is translated into an object graph navigation with the traversal root being the selected book. After looking at the related information for the selected book, the user may choose to look at another book by the same author. When information relating to the newly selected book is requested, the newly selected book becomes the root of a new object graph traversal.

Next, we give an overview of the five main steps of the dynamic framework and in the process show how the above example scenario fits in.

1. H-region parameters specification: The dynamic framework divides the database into regions of homogeneous access probability (H-regions). In our example, each H-region represents a different group of books, each group having its own probability of access. In this step, we specify the characteristics of each H-region, e.g., its size, initial access probability, etc.

2. Workload specification. H-regions are responsible for assigning access probability to pieces of data (tuples or objects). However, H-regions do not dictate what to do then. We term the selected tuple or object workload root. In the remainder of this paper we will use the term "root" to mean workload root. In this step, we select the type of workload to execute after selecting the root.

3. Regional protocol specification. Regional protocols use H-regions to accomplish access pattern change. Different styles of access pattern change can be accomplished by changing the H-region parameter values with time. For example, a regional protocol may initially define one H-region with a high access probability, while the remaining H-regions are assigned low access probabilities. After a certain time interval, a different H-region may become the high access probability region. This, when translated to the book store example, is similar to Australian travel books becoming less popular after the 2000 Olympics end.

4. Dependency protocol specification. Dependency protocols allow us to specify a relationship between the currently selected root and the next root. In our example, this is reflected in the customer deciding to select a book which is by the same author as the previously selected book.

5. Regional and dependency protocol integration specification. In this step, regional and dependency protocols are integrated to model changes in dependency between successive roots. An example is a customer using our on-line book store, who selects a book of interest, and then is confronted with a list of currently popular books by the same author. The customer then selects one of the listed books (modeled by dependency protocol). The set of currently popular books by the same author may change with time (modeled by regional protocol).

The first three steps we have described are generic, i.e., they can be applied on any selected benchmark and system type (relational, object-oriented, or object-relational). The two last steps are similar when varying the system type, but are nonetheless different because access paths and

Methods are substantially different in a relational system (with tables, tuples, and joins) and an object-oriented system (with objects and references), for instance.

Next, we further detail the concept of H-region and the generic regional protocol specification.

## H-regions

H-regions are created by partitioning the objects of the database into non-overlaping sets. All objects in the same H-region has the same access probability. Here we use the term access probability to mean the likelihood that an individual object of the H-region will be accessed at a given moment in time. The parameters that define an H-region are listed below.

- HR-SIZE: The size of the H-region is specified as a fraction of the database size. Constraint: The sum size of all regions must equal 1.
- INIT-PROB-W: The initial probability weight that is assigned to the region. The actual probability is derived from the probability weight, by dividing the probability weight of the region by the sum probability weight of all regions.
- LOWEST-PROB-W: The lowest probability weight this region can go down to.
- HIGHEST-PROB-W: The highest probability weight this region can go up to.
- PROB-W-INCR-SIZE: The amount by which the probability weight of this region increases or decreases when change is requested.
- OBJECT-ASSIGN-METHOD: This determines the way objects are assigned into this region. The options are random selection and by class selection. Random selection picks objects randomly from anywhere in the database. By class selection places attempts to assign objects of the same class into the same H-region, as much as possible. It first sorts objects by class ID and then picks the first $N$ objects (in sorted order), where $N$ is the number of objects allocated to the H-region.
- INIT-DIR: The initial direction that the probability weight increment moves in.

The access probability of an H-region can never be below LOWEST-PROB-W or above HIGHEST-PROB-W.

### Regional Protocols

Regional protocols simulate access pattern change by first initializing the parameters of every H-region, and then periodically changing the parameter values in certain predefined ways. This paper documents three styles of regional change: moving window of change, gradual moving window of change, and cycles of change. Although these three styles of change together provide a good spectrum of ways in which access pattern can change, they are by no means exhaustive. Other researchers or framework users are encouraged to create new regional protocols of their own.

**Moving Window of Change Protocol.** This regional protocol simulates sudden changes in access pattern. In our on-line book store, this is translated to books suddenly becoming popular due to some event, and once the event passes, the books become unpopular very fast. For instance, books that are recommended in a TV show may become very popular in the few days after the show, but may quickly become unpopular when the next set of books are introduced. This style of change is accomplished by moving a window through the database. The objects in the window have a much higher probability of being chosen as root when compared to the remainder of the database. This is done by breaking up the database into $N$ H-regions of equal size. One H-region is first initialised to be the hot region (where heat is used to denote probability of reference), and then after $H$ root selections, a different H-region becomes the hot region. $H$ is a user-defined parameter that reflects the rate of access pattern change.

- The database is broken up into $N$ regions of equal size.
- All H-regions have the same value for HIGHEST-PROB-W, LOWEST-PROB-W and PROB-W-INCR-SIZE.
- Set the INIT-PROB-W of one of the H-regions to equal HIGHES-PROB-W (the hot region) and the rest of the H-regions get their INIT-PROB-W assigned to LOWEST-PROB-W.
- Set PROB-W-INCR-SIZE of every region to equal HIGHEST-PROB-W - LOWEST-PROB-W.
- The INIT-DIR parameter of all the H-regions are set to move downwards. Initially, the window is placed at the hot region. After every $H$ root selections, the window moves from one H-region to another. The H-region that the window is moving from has its direction set to down. The H-region that the window is moving into has its direction set to up. Then, probability weights of the H-regions are incremented or decremented depending on the current direction of movement.

**Gradual Moving Window of Change Protocol.** The way this protocol differs from the previous one is that the hot region cools down gradually instead of suddenly. The cold regions also heat up gradually as the window is moved onto them. In our book store example, this style of change may depict travel guides to Australia gradually becoming less popular after the Sydney 2000 Olympics. As a consequence, travel guides to other countries may gradually become more popular. Gradual changes of heat may be more common in the real world.

This protocol is specified in the same way as the previous protocol with two exceptions. First, PROB-W-INCR-SIZE is now user-specified instead of being the difference between HIGHEST-PROB-W and LOWEST-PROB-W. The value of PROB-W-INCR-SIZE determines how vigorously access pattern changes at every change iteration. We use the term change iteration to mean the changing of access probabilities of the H-regions after every H (defined in the previous section) root selections. The second exception is in the way the H-regions change direction. The H-region that the window moves into has its direction toggled. The direction of the H-region that the window is moving from is unchanged. This way, the previous H-region is able to continue cooling down gradually or heating up gradually. When the access probability of a cooling H-region reaches its LOWEST-PROB-W it stops cooling and similarly a heating up H-region stops heating up when it reaches its HIGHEST-PROB-W.

**Cycles of Change Protocol.** This style of change mimics something like a bank where customers in the morning tend to be of one type (e.g., social category), and in the afternoon of another type. This, when repeated, creates a cycle of change. Cycles of change can be simulated using the following steps. Members of a set are not ordered.

- Break up the database into three H-regions. The first two H-regions represent objects going through the cycle of change. The third H-region represents the remaining unchanged part of the database.

- The HR-SIZE of the first two H-regions are equal to each other and user-specified. The HR-SIZE of the third H-region is equal to the remaining fraction of the database.
- Set the LOWEST-PROB-W and HIGHEST-PROB-W parameters of the first two H-regions to values that reflect the two extremes of the cycle.
- Set the PROB-W-INCR-SIZE of the first two H-regions to both equal HIGHEST-PROB-W - LOWEST-PROB-W. Set the PROB-W-INCR-SIZE of the third H-region to equal zero.
- The INIT-PROB-W of the first H-region is set to HIGHEST-PROB-W and the second to LOWEST-PROB-W.
- Set the INIT-DIR of the hot H-region to down and the INIT-DIR of the cold H-region to up.
- Again, the H parameter is used to vary the rate of access pattern change.

## The Dynamic Object Evaluation Framework (DoEF)

In this section we describe DoEF, which is an instance of DEF. In DoEF, the workload type is selected from those defined in OCB. For sequential scans, the class of the root object is used to decide which objects are scanned; for traversals, the root object becomes the root of the traversal; and for updates, either the class of the root object or just the root object is used to decide which objects are updated (depending on the particular update workload selected).

Next, we detail the steps in DEF that are specific to the object-oriented context, i.e., the specification of the depedency protocols and their integration with the regional protocols.

### Dependency Protocols

There are many scenarios in which a person executes a query and then decides to execute another query based on the results of the first query, thus establishing a dependency between the two queries. In this paper, we have specified four dependency protocols: random selection protocol, by reference selection protocol, traversed objects selection protocol, and same class selection protocol. Again, these protocols are not meant to be exhaustive and other researchers or benchmark users are encouraged to extend DoEF beyond these dependency protocols.

**Random Selection Protocol.** This method simply uses some random function to select the current root. This protocol mimics a person starting a completely new query after finishing the previous one.

$r_i = RAND1()$

$r_i$ is the ID of the $i^{th}$ root object. The function $RAND1()$ can be any random function. An example of $RAND1()$ is a skewed random function that selects a certain group of root objects with a higher probability than others.

**By Reference Selection Protocol.** The current root is chosen to be an object referenced by the previous root. An example of this protocol in our on-line book store scenario is a person having finished with a selected book, who then decides to look at the next book in the series (assuming the books of the same series are linked together by structural references).

$r_{i+1} = RAND2(RefSet(r_i, D))$

$RefSet(r_i, D)$ is a function that returns the set of objects that the $i^{th}$ root references. $RAND2()$, like $RAND1()$ can be any random function. Two types of references can be used: structure references (S-references) and D-references. Structure references are simply the references obtained from the object graph. D-references are a new type of reference used for the sole purpose of establishing dependencies between roots of traversals. The parameter D is used to specify the number of D-references per object. Note if structure references are specified, then parameter D is not used.

**Traversed Objects Selection Protocol.** The current root is selected from the set of objects that are referenced in the previous traversal. An example is a customer in the first query requesting a list of books along with their authors and publishers (thus requiring the book objects themselves to be retrieved), who then decides to read an exerpt from one of the books listed.

$r_{i+1} = RAND3(TraversedSet(r_i, C))$

$TraversedSet(r_i, C)$ returns the set of objects referenced during the traversal that began with the $i^{th}$ root. $RAND3()$, like $RAND1()$ can be any random function. The parameter C is used to restrict the number of objects returned by $TraversedSet(r_i, C)$. C is specified as a fraction of the objects traversed. This way, the degree of locality of objects returned by $TraversedSet(r_i, C)$ can be controlled (smaller C means higher degree of locality).

**Same Class Selection Protocol.** In same class selection, the currently selected root must belong to the same class as the previous root. Root selection is further restricted to a subset of objects of the class. The subset is chosen by a function that takes the previous root as a parameter. That is, the subset chosen is dependent on the previous root object. An example of this protocol is a customer deciding to select a book from our on-line book store which is by the same author as the previous selected book. In this case, the same class selection function returns books by the same author as the selected book.

$$r_{i+1} = RAND4(f(r_i, Class(r_i), U))$$

$Class(r_i)$ returns the class of the $i^{th}$ root. $RAND4()$, like $RAND1()$ can be any random function. The parameter U is user-defined and specifies the size of the set returned by function $f()$. U is specified as a fraction of the total class size. U can be used to increase or decrease the degree of locality between the objects returned by $f()$. $f()$ always returns the same set of objects given the same set of parameters.

**Hybrid Setting.** The hybrid setting allows an experiment to use a mixture of the dependency protocols outlined above. Its use is important since it simulates a user starting a fresh random query after having followed a few dependencies. Thus, the hybrid setting is implemented in two phases. The first randomisation phase uses the random selection protocol to randomly select a root. In the second dependency phase, one of the dependency protocols outlined in the previous section is used to select the next root. R iterations of the second phase are repeated before going back to the first phase. The two phases are repeated continuously.

The probability of selecting a particular dependency protocol during the dependency phase is specified via the following settings: RANDOM-DEP-PROB (random selection), SREF-DEP-PROB (by reference selection using structure references), DREF-DEP-PROB (by reference selection using D-references), TRAVERSED-DEP-PROB (traversed objects selection), and CLASS-DEP-PROB (same class selection).

## Integration of Regional and Dependency Protocols

Dependency protocols model user behavior. Since user behavior can change with time, dependency protocols should also be able to change with time. The integration of regional and dependency protocols allows us to simulate changes in the dependency between successive root selections. This is easily accomplished by exploiting the dependency protocols' property of returning a candidate set of objects when given a particular previous root. Up to now, the next root is selected from the candidate set by the use of a random function. Instead of using the random function, we partition the candidate set using H-regions and then apply regional protocols on these H-regions. When integrating with the traversed objects dependency protocol, the following property must hold: whenever given the same root object, the same set of objects is always traversed. This way, the same previous root will return the same candidate set.

## Tested Systems and Algorithms

In this section, we briefly describe the dynamic clustering algorithms and object stores we have used in the experimental section of this paper.

## Dynamic Clustering Algorithms

Dynamic clustering is the periodic on-line re-organisation of objects in an ODBMS. The aim is to allow the physical placement of objects on disk to more closely reflect the pervading pattern of database access. Objects that are likely to be accessed together in the near future are placed in the same page, thereby reducing the number of disk I/Os.

Dynamic, Statistical and Tunable Clustering (DSTC). DSTC (Bullat and Schneider, 1996) is a dynamic clustering algorithm that has the feature of achieving dynamicity without adding high statistics collection overhead and excessive volume of statistics. However, it does not take care to reduce I/O generated from the clustering process itself.

The clustering algorithm is not very selective when deciding which pages to re-cluster. The effect is a page is re-clustered even if there is a slight benefit in re-clustering it. However, the slight benefit gained from re-clustering is often outweighed by the cost of loading the page into memory for re-clustering. This situation (re-clustering of slightly badly clustered pages) will become more frequent as access pattern changes more rapidly. It is for this reason we expect that DSTC will perform poorly when access pattern changes rapidly.

**Detection & Reclustering of Objects (DRO).** Learning from the experiences of DSTC and StatClust (Gay and Gruenwald, 1997), DRO (Darmont et al., 2000) is designed to produce less clustering I/O overhead and use less statistics. DRO uses various thresholds to limit the pages involved in re-clustering to only the pages that are most in need of re-clustering. We term this flexible conservative re-clustering. Experiments conducted using OCB show that DRO outperforms DSTC (Darmont et al., 2000). The improvement in performance is mainly attributed to the low clustering I/O overhead of DRO. In order to limit statistics collection overhead, DRO only uses object frequency and page usage rate information. In contrast, DSTC stores object transition information, which is much more costly. Since DRO chooses only a limited number of the worst clustered pages to re-cluster (flexible conservative re-clustering) it should perform better than DSTC when access pattern changes rapidly. This is because when access pattern changes rapidly, the benefits in re-clustering pages become lower and thus there will be more pages which only benefit slightly from re-clustering. DRO does not re-cluster these pages whereas DSTC does. This leads DSTC to generate larger clustering overhead for very slight improvements in clustering quality.

**Opportunistic Prioritised Clustering Framework (OPCF).** OPCF (He et al., 2000) is a framework for translating any static clustering algorithm (where re-clustering occurs off-line) into a dynamic clustering algorithm. OPCF creates algorithms that have the following key properties: read and write I/O opportunism and prioritization of re-clustering. Read and write I/O opportunism refers to limiting re-clustering to pages that are currently in memory (in the case of read opportunism) and dirty (in the case of write opportunism). This approach reduces the I/O overhead associated with re-clustering. Prioritization of re-clustering refers to choosing a limited number of the worst clustered pages to be re-clustered first. This also reduces clustering overhead by reducing the number of pages re-clustered. Therefore OPCF clustering algorithms also perform flexible conservative re-clustering. Two dynamic clustering algorithms produced from the OPCF framework are presented in (He et al., 2000): dynamic graph partitioning algorithm (GP) and dynamic probability ranking principle algorithm (PRP).

Since OPCF, like DRO, performs flexible conservative re-clustering, it should also perform well when access pattern changes very rapidly. We will use the term flexible clustering algorithms to refer to DRO and the OPCF dynamic clustering algorithms.

**Analysis of dynamic clustering algorithms.** In this section we analyze the relative performances of the dynamic clustering algorithms based on the characteristics of the algorithms. For the moving window of change protocol we expect the relative difference in performance between DSTC and the flexible clustering algorithms to increase with increasing rate of change. This is because DSTC does not do flexible conservative re-clustering and thus incurs high re-clustering overheads.

The relative difference between the different flexible clustering algorithms should not change by much with increasing rate of change since they are all limit the clustering overheads to a bounded amount. In terms of the shapes of the curves, we expect DSTC to perform linearly worse with increasing rate of change. This is because it does not bound the clustering overhead. In contrast the flexible dynamic clustering algorithms' performance will increase with increasing rate of change but flat after a certain point (we call this the saturation point). This is because these algorithms bound the clustering overhead.

In terms of the gradual moving window of change protocol, we expect the relative differences between the algorithms to stay similar as the rate of change increases. The reason is this change protocol is very mild and therefore do not cause the flexible clustering algorithms to reach their saturation point. In terms of the shapes of the curves, we expect the performance of all the algorithms to perform close to linear with increasing rate of change of access pattern. This is because increases in the rate of change of access pattern causes the benefit of re-clustering to diminish, this increase is constant and does not reach a saturation point due to the mild style of change.

**Object stores**

**Platypus.** Platypus (He et al., 2000) is a flexible high performance transactional object store, designed to be used as the storage manager for persistent programming languages. The design includes support for SMP concurrency: stand-alone, client-server and client-peer distribution configurations; configurable logging and recovery; and object management that can accommodate garbage collection and clustering mechanisms. In addition to these features, Platypus is built for speed. It features a new recovery algorithm derived from the popular ARIES (Mohan, Haderle, Lindsay, Pirahesh, and Schwarz, 1992) recovery algorithm, which removes the need for log sequence numbers to be present in store pages; a zero-copy memory-mapped buffer manager with controlled write-back behaviour; and a novel fast and scalable data structure (splay trees) used extensively for accessing metadata.

**SHORE.** SHORE (Carey et al., 1994) is a transactional persistent object system that is designed to serve the needs of a wide variety of target applications, including persistent programming languages. It has a peer-to-peer distribution configuration. Like Platypus, it also has a focus on performance.

**Analysis of dynamic object stores.** In this section we analyze the relative performances of Platypus and SHORE. Since they handle swapping differently we expect Platypus to show better performance than SHORE when rate of access pattern change is small but its lead diminishes as the rate of access pattern change increases. The reason lies in the change in access locality when the rate of access pattern change. When the rate of access pattern change is low, access locality is high (due to small and slow moving hot region), and thus most object requests can be satisfied from the buffer cache. However as the rate of access pattern change increases, access locality diminishes which results in more buffer cache misses. Thus, the reason behind Platypus' poor performance lies in its poor swapping performance. Platypus's poor swapping performance is due to the low degree of concurrency (coarse grained locking) between the page server and the client process when swapping is in progress (a deficiency in the implementation).

## Experimental Results

This section details two sets of experiments we have conducted to evaluate the effectiveness of DoEF. In the first set of experiments four state of the art dynamic clustering algorithms are benchmarked. In the second set two real object stores are benchmarked.

For dynamic clustering algorithms we have conducted two sets of experiments: moving and gradual moving window of change regional protocol experiments; and moving and gradual moving S-reference protocol experiments. For the real object stores, we also conducted two sets of experiments: moving window of change protocol experiments; and moving window of change traversed objects experiment. There are two reasons for choosing these set of protocols to test: the space constraints prohibit us from showing results obtained using all combinations of protocols; and after testing many of the possible combinations we found for the particular clustering algorithms and real OODBs we have tested, the experiments presented gives the greatest insight into the effectiveness of DoEF.

### Dynamic Clustering Experiments

These experiments use DoEF to compare the performance of four state of the art dynamic clustering algorithms: DSTC, DRO, OPCF-PRP, and OPCF-GP. The parameters we have used for the dynamic clustering algorithms are shown in Table 2. In the interests of space, we do not include their description in this paper. However, they are wholly described in their respective papers. The clustering techniques have been parameterized for the same behavior and best performance.

| Parameter | Value |
|---|---|
| n | 200 |
| $n_p$ | 1 |
| p | 1000 |
| $T_{fa}$ | 1.0 |
| $T_{fe}$ | 1.0 |
| $T_{fc}$ | 1.0 |
| w | 0.3 |

(a) DSTC

| Parameter | Value |
|---|---|
| MinUR | 0.001 |
| MinLT | 2 |
| PCRate | 0.02 |
| MaxD | 1 |
| MaxDR | 0.2 |
| MaxRR | 0.95 |
| SUInd | true |

(b) DRO

| Parameter | PRP value | GP value |
|---|---|---|
| N | 200 | 200 |
| CBT | 0.1 | 0.1 |
| NPA | 50 | 50 |
| NRI | 25 | 25 |

(c) OPCF

Table 2. DSTC, DRO, OPCF-PRP, and OPCF-GP parameters

The experiments are conducted on the Virtual Object-Oriented Database simulator (VOODB) (Darmont and Schneider, 1999). VOODB is based on a generic discrete-event simulation framework. Its purpose is to allow performance evaluations of OODBMSs in general, and optimisation methods like clustering in particular. VOODB has been validated for two real-world OODBMSs, $O_2$ (Deux, 1991) and Texas (Singhal, Kakkad, and Wilson, 1992). The VOODB parameter values we have used are depicted in Table 3 (a). Simulation is chosen for this experiment for two reasons. First, it allows rapid development and testing of a large number of dynamic clustering algorithms (all previous dynamic clustering papers compared at most two algorithms). Second, it is relatively easy to simulate accurately, read, write and clustering I/O (the dominating metrics that determine the performance of dynamic clustering algorithms).

Since DoEF uses the OCB database and operations, it is important for us to document the OCB settings we have used for these experiments. The values of the database parameters we have used are shown in Table 3 (b). The sizes of the objects we have used varies from 50 to 1600 bytes, with the average size being 233 bytes. A total of 100,000 objects are generated for a total database size of 23.3 MB. Although this is a small database size, we have also used a small buffer size (4 MB) to keep the database to buffer size ratio large. Clustering algorithm performance is indeed more sensitive to database to buffer size ratio than database size alone. The operation we have used for all the experiments is the simple, depth-first traversal with traversal depth 2. The simple traversal is chosen since it is the only traversal that always accesses the same set of objects given a particular root. This establishes a direct relationship between varying root selection and changes in access pattern. Each experiment involved executing 10,000 transactions.

The main DoEF parameter settings we have used in this study are shown in Table 4. These DoEF settings are common to all experiments in this paper. The HR-SIZE setting of 0.003 (remember this is the database population from which the traversal root is selected) creates a hot region about 3% the size of the database (each traversal touches approximately 10 objects). This fact is verified from statistical analysis of the trace generated. The HIGHEST-PROB-W setting of 0.8 and LOWEST-PROB-W setting of 0.0006, produces a hot region with 80% probability of reference and the remaining cold regions with a combined reference probability of 20%. These settings are chosen to represent typical database application behaviour. Gray cites statistics from a real videotext application in which 3% of the records got 80% of the references (Gray and Putzolu, 1987). Carey uses a hot region size of 4% with a 80% probability of being referenced in the HOTCOLD workload we have used to measure data caching tradeoffs in client-server OODBMSs (Carey et al., 1991). Franklin uses a hot region size of 2% with a 80% probability of being referenced in the HOTCOLD workload we have used to measure the effects of local disk caching for client server OODBMSs (Franklin et al., 1993). In addition to the results reported in this paper, we also tested the sensitivity of the results to variations in hot region size and probability of reference. We found the algorithms show similar general tendencies at different hot region sizes and probability of reference. It is for this reason and in the interests of space we omit these results.

The dynamic clustering algorithms shown on the graphs in this section are labeled as follows:

| Parameter description | Value |
|---|---|
| System Class | Centralized |
| Disk page size | 4096 bytes |
| Buffer size | 4 MB |
| Buffer replacement policy | LRU-1 |
| Pre-fetching policy | None |
| Multiprogramming level | 1 |
| Number of users | 1 |
| Object initial placement | Sequential |

(a) VOODB parameters

| Parameter description | Value |
|---|---|
| Number of classes in the database | 50 |
| Maximum number of references, per class | 10 |
| Instances base size, per class | 50 |
| Total number of objects | 100000 |
| Number of reference types | 4 |
| Reference types random distribution | Uniform |
| Class reference random distribution | Uniform |
| Objects in classes random distribution | Uniform |
| Objects references random distribution | Uniform |

(b) OCB parameters

Table 3. VOODB and OCB parameters

| Parameter name | Value |
|---|---|
| HR-SIZE | 0.003 |
| HIGHEST-PROB-W | 0.80 |
| LOWEST-PROB-W | 0.0006 |
| PROB-W-INCR-SIZE | 0.02 |
| OBJECT-ASSIGN-METHOD | Random object assignment |

Table 4. DoEF parameters

- NC: No Clustering;

- DSTC: Dynamic Statistical Tunable Clustering;

- GP: OPCF (greedy graph partitioning);

- PRP: OPCF (probability ranking principle);

- DRO: Detection and Re-clustering of Objects.

As we discuss the results of these experiments, we focus our discussion on the relative ability of each algorithm to adapt to changes in access pattern, i.e., as rate of access pattern change increases, we seek to know which algorithm exhibits more rapid performance deterioration. This contrasts from discussing which algorithm gives the best absolute performance. All the results presented here are in terms of total I/O. Total I/O is the sum of transaction read I/O, clustering read and clustering write I/O. Thus, the results give an overall indication of the performance of each clustering algorithm, including each algorithm's clustering I/O overhead.

**Moving and Gradual Moving Regional Experiments.** In these experiments, we have used the regional protocols moving window of change and gradual moving window of change to test each of the dynamic clustering algorithms' ability to adapt to changes in access pattern. The regional protocol settings we have used are shown in Table 4. We vary the parameter $H$, rate of access pattern change. The results for these experiments are shown in Figure 2. There are three main results from this experiment. Firstly, when rate of access pattern change is small (when parameter H is less than 0.0006 in Figure 2 (a) and all of Figure 2 (b)), all algorithms show similar performance trends (rate of performance degradation). This implies at moderate levels of access pattern change all algorithms are approximately equal in their ability to adapt to the change. Secondly, when the more vigorous style of change is applied (Figure 2 (a)), all dynamic clustering algorithms' performance quickly degrades to worse than no clustering. Thirdly, when access pattern change is very vigorous (when paramter H is greater than 0.0006 in Figure 2 (a)), DRO and OPCF algorithms GP and PRP show a better trend performance (rate of performance degradation), implying DRO and OPCF are more robust to access pattern change. This supports our analysis described in analysis of dynamic clustering algorithms section.

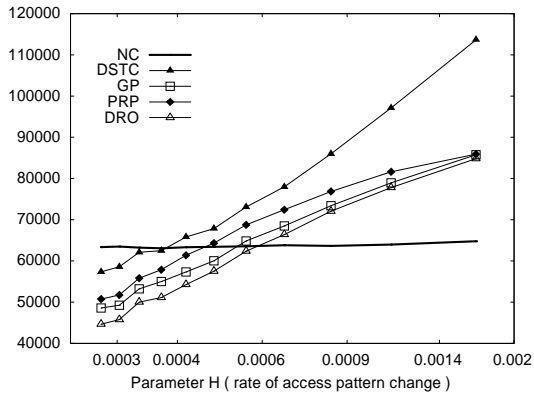 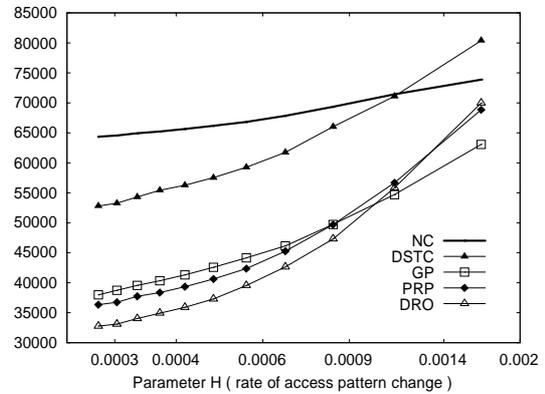

(a) Moving Window of Change        (b) Gradual Moving Window of Change

Fig. 2. Regional dependency results. The x-axis is in $\log_2$ scale.

**Moving and Gradual Moving S-Reference Experiments.** In these experiments, we explore the effect that changing pattern of access has on the S-reference dependency protocol. This is accomplished by using the integrated regional dependency protocol method. We integrated S-reference dependency with the moving and gradual moving window of change regional protocols. For this experiment, we use the hybrid dependency setting. R is set to 1. The first phase (random phase) of the hybrid setting requires a random dependency function. The random function we use partitions the database into one hot and one cold region. The hot region is set to be 3% of the database size and has an 80% probability of reference (typical database application behavior (Gray and Putzolu, 1987; Carey et al., 1991; Franklin et al., 1993)). S-reference dependency is the only dependency protocol used. The regional protocol settings are as described in Table 4.

The results for these experiments are shown in Figure 3. In the moving window of change results (Figure 3 (a)), DRO and the OPCF algorithms (GP and PRP) are again more robust to changes in access pattern than DSTC for moving window of change. However, in contrast to the previous experiment, DRO and OPCF algorithms never perform worse than NC by much, even when parameter H is 1 (access pattern changes after every transaction). The reason is that the cooling and heating of S-references is a milder form of access pattern change than the pure moving window of change regional protocol of the previous experiment. In the gradual moving window of change results shown in figure 3 (b), all dynamic clustering algorithms show approximately the same trend performance. This is similar to the observation made in the previous experiment. This supports our analysis described in analysis of dynamic clustering algorithms section.

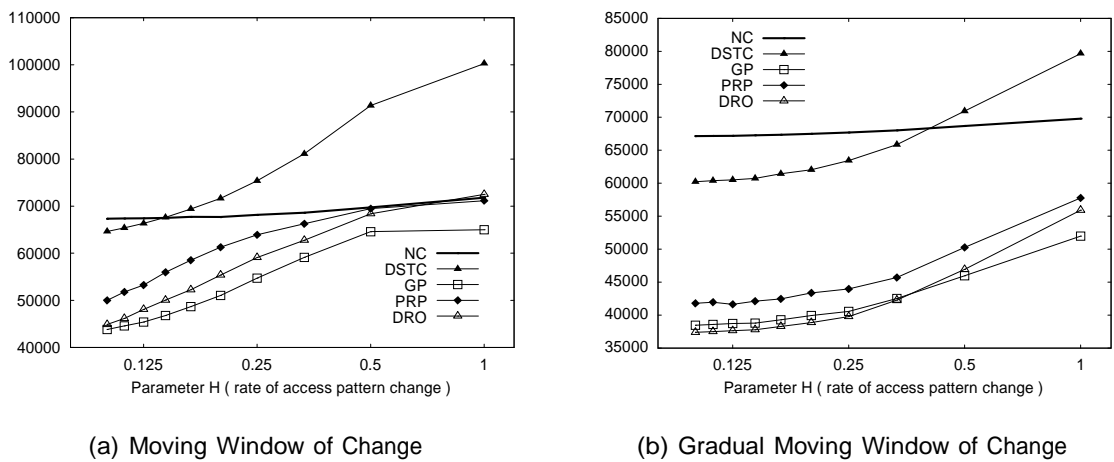

(a) Moving Window of Change     (b) Gradual Moving Window of Change

Fig. 3. S-reference dependency results. The x-axis is in $\log_2$ scale.

## Object Store Experiments

In this section, we report the results of using DoEF to compare the performance of two real object stores: SHORE and Platypus.

SHORE has a layered architecture that allows users to choose the level of support appropriate for a particular application. The lowest layer of SHORE is the SHORE Storage Manager (SSM), which provides basic object reading and writing primitives. Using the SHORE SSM, we have constructed PSI-SSM, a thin layer providing PSI (Blackburn, 1998) compliance for SSM. By using the PSI interface, the same DoEF code could be used for both Platypus and SHORE. The buffer replacement policy that SHORE uses is CLOCK. We use SHORE version 2.0 for all of the experiments reported in this paper.

The Platypus implementation we have used for this set of experiments has the following features: physical object IDs; the NOFORCE/STEAL recovery policy (Franklin, 1997); zero-copy memory mapped buffer cache; the use of hash-splay trees to manage metadata; PSI compliance; the system is SMP re-entrant and supports multiple concurrent client threads per process as well as multiple concurrent client processes. Limitations of the Platypus implementation at the time of writing include: the failure-recovery process is untested (although logging is fully implemented); virtual address space reserved for metadata is not recycled; the store lacks a sophisticated object manager with support for garbage collection and dynamic clustering, and lacks support for distributed store configurations. Platypus uses the LRU replacement policy.

In this sets of experiments, the SHORE and Platypus implementations do not include dynamic clustering algorithms. In contrast to the previous experiment, we are interested here in comparing the other factors (besides clustering) that affect system performance. The experiments in this section are conducted using Solaris 7 on an Intel machine with dual Celeron 433Mhz processors, 512 MB of memory and a 4 GB hard disk. The OCB database and workload settings we have used for this experiment are the same as for the previous set of experiments, with the exception that a total of 400,000 objects is generated instead of 100,000. The reason for using a larger database size is that the real object stores are configured with a larger buffer cache, therefore we need to increase the database size in order to test the swapping. The size of the objects we have used vary from 50 to 1200 bytes, with the average size being 269 bytes. Therefore, the total database size is 108 MB.

**Moving Window of Change Regional Experiment.** In this experiment, we use the moving window of change protocol to compare the effects that changing pattern of access has on Platypus and SHORE. The regional protocol settings we have used are the same as shown in Table 4. The buffer size is set to 61 MB. Note that both Platypus and SHORE have there own buffer managers with user-defined buffer sizes.

The results for this experiment are shown in Figure 4. The results show the trend predicted in the analysis of object stores section, namely the performance of Platypus start well in front of SHORE but its lead rapidly diminishes as the rate of access pattern change increases.

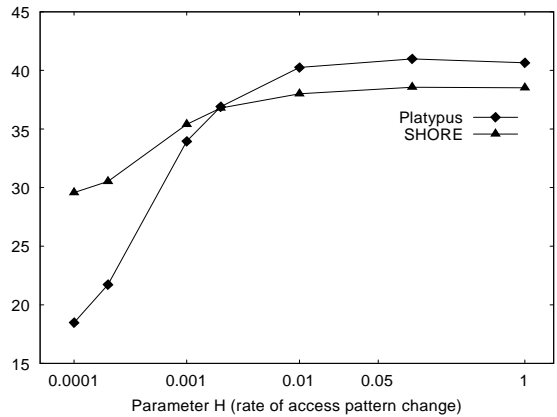

Fig. 4. Moving window of change regional protocol results. The x-axis is in $\log_2$ scale.

**Moving Window of Change Traversed Objects Experiment.** In this experiment, we compare the performance of Platypus and SHORE in the context of moving traversed objects dependency protocol. This is accomplished by using the integrated regional dependency protocol method outlined in. We have integrated traversed objects dependency protocol with the moving window regional protocol. For this experiment, we use the hybrid dependency setting. R is set to 1. The random function we use partitions the database into one hot and one cold region. The hot region is set to be 0.01 fraction of the database size, and the cold region is assigned the remaining portion of the database. 99% of the roots are selected from the hot region. The C parameter is set to 1.0. Traversed objects dependency is the only dependency protocol we have used. The regional protocol parameters we have used are identical to those used in the previous experiment, with the exception HR SIZE is set to 0.05. In this experiment, the buffer size we have used is only 20 MB as opposed to 61 MB in the previous experiment, because this experiment has a smaller working set size, thus, at 61 MB, swapping would not occur (even when H is one). The reason behind the small working set size lies in the fact that the random function we have used does not move its hot region. The results for this experiment are shown in Figure 5. The results again show the performance of Platypus diminishes at a faster rate than SHORE. The reason for this behavior can again be explained by Platypus' poor swapping performance. However, the saturation occurs later than for the moving window of change protocol since the degree of locality in this protocol is higher.

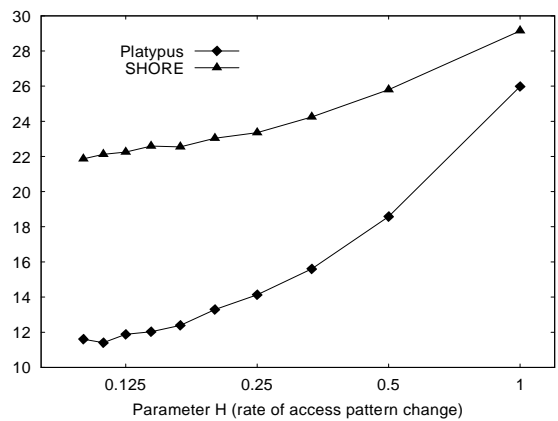

Fig. 5. Traversed objects results. The x-axis is in $\log_2$ scale. The minimum and maximum coefficients of variation are 0.005 and 0.037, respectively.

## Conclusion

In this paper, we have detailed the specification of a new framework for database benchmarking, DEF, which allows DBMSs' designers and users to test the performances of a given system in a dynamic setting. This is where the originality of our work lies, since almost all real world applications exhibit access pattern changes, but no existing benchmark attempts to model this behavior. We have also instantiated DEF in an object-oriented context under the name of DoEF, to illustrate how such a specialization could be performed.

We have designed DEF to be readily extensible along two axes. First, since this is, to the best of our knowledge, the first attempt at studying the dynamic behavior of DBMSs, we have taken great care to make the incorporation of new styles of access pattern change as painless as possible, mainly through the definition of H-regions. We actually view the DEF software as an open platform that is available to other researchers for testing their own ideas. The DoEF code we have used in both our object clustering simulation experiments and our implementation for Platypus and SHORE is freely available for download.

Second, although we have considered an object-oriented environment in this study with DoEF, we can also apply the concepts developed in this paper to other types of databases. Instantiating DEF for object-relational databases, for instance, should be relatively easy. Since OCB can be quite easily adapted to the object-relational context (even if extensions would be required, such as abstract data types or nested tables, for instance), DEF can be used in the object-relational context too.

As stated in the introduction, the main objective of DEF is to allow researchers and engineers to explore the performance of databases (identify components that are causing poor performance) within the context of changing patterns of data access. Our experimental results involving dynamic clustering algorithms and real object stores have indeed demonstrated DoEF's ability to meet this objective. Within the dynamic clustering context, two hypothesis have been confirmed by our experiments: (1) dynamic clustering algorithms can deal with moderate levels of access pattern change but performance rapidly degrades to be worse than no clustering when vigorous styles of access pattern change are applied; and (2) flexible conservative re-clustering is the key in determining a clustering algorithm's ability to adapt to changes in access pattern. In the performance comparison between the real object stores Platypus and SHORE, the use of DoEF allowed us to identify Platypus' poor swapping performance.

This study opens several research perspectives. The first one concerns the exploitation of DEF to keep on acquiring knowledge about the dynamic behavior of various DBMSs. Furthermore, comparing the dynamic behavior of different systems, though an interesting task in itself, may inspire us to develop new styles of access pattern change. New styles of access pattern change identified in this and other ways may be incorporated into DEF.

Finally, the effectiveness of DEF at evaluating other aspects of database performance can also be explored. Data clustering is indeed an important performance optimization technique, but other strategies such as buffer replacement and pre-fetching should also be evaluated.

## Acknowledgements

The authors would like to thank Stephen M. Blackburn for his helpful comments and suggestions related to the writing of this paper. In addition his help in making SHORE and Platypus available for benchmarking was invaluable. We also thank the anonymous reviewers for their detailed and constructive feedback.

# References

Anderson, T., Berre, A., Mallison, M., Porter, H., & Scheider, B. (1990). The HyperModel benchmark. In International Conference on Extending Database Technology, pp. 317–331.

Blackburn, S. M. (1998). Persistent Store Interface: A foundation for scalable persistent system design. Ph.D. thesis, Australian National University, Canberra, Australia.

Bullat, F., & Schneider, M. (1996). Dynamic clustering in object databases exploiting effective use of relationships between objects. In ECOOP'96, 10th European Conference on Object-Oriented Programming, July 8-12, 1996, Linz, Austria, pp. 344–365. Springer.

Carey, M., DeWitt, D., & Naughton, J. (1993). The OO7 benchmark. ACM-SIGMOD International Conference on the Management of Data (SIGMOD 93).

Carey, M., DeWitt, D., Naughton, J., Asgarian, M., Brown, P., Gehrke, J., & Shah, D. (1997). The bucky object-relational benchmark. In ACM SIGMOD International Conference on Management of Data (SIGMOD 97), pp. 135–146 Tucson, Arizona, USA.

Carey, M. J., DeWitt, D. J., Franklin, M. J., Hall, N. E., McAuliffe, M., Naughton, J. F., Schuh, D. T., Solomon, M. H., Tan, C. K., Tsatalos, O., White, S., & Zwilling, M. J. (1994). Shoring up persistent applications. In ACM-SIGMOD Conference on the Management of Data Minneapolis.

Carey, M. J., Franklin, M. J., Livny, M., & Shekita, E. J. (1991). Data caching tradeoffs in client-server DBMS architectures. In Clifford, J., & King, R. (Eds.), In Proceedings of the ACM SIGMOD conference on Management of Data, pp. 357–366.

Cattell, R. (1991). An engineering database benchmark. In Gray, J. (Ed.), The Benchmark Handbook for Database Transaction Processing Systems, pp. 247–281. Morgan Kaufmann.

Darmont, J., Fromantin, C., Regnier, S., Gruenwald, L., & Schneider, M. (2000). Dynamic clustering in object-oriented databases: An advocacy for simplicity. In International Symposium on Object and Databases, Vol. 1944 of LNCS, pp. 71–85.

Darmont, J., Petit, B., & Schneider, M. (1998). OCB: A generic benchmark to evaluate the per- formances of object-oriented database systems. In 6th International Conference on Extending Database Technology (EDBT '98), Vol. 1377 of LNCS, pp. 326–340.

Darmont, J., & Schneider, M. (1999). VOODB: A generic discrete-event random simulation model to evaluate the performances of OODBs. In 25th VLDB conference, Edinburgh, Scotland, pp. 254–265.

Darmont, J., & Schneider, M. (2000). Benchmarking OODBs with a generic tool. Journal of Database Management, 11(3), 16–27.

Deux, O. (1991). The $O_2$ system. Communications of ACM, 34(10), 34–48.

Franklin, M. J. (1997). Concurrency control and recovery. In Tucker, A. B. (Ed.), The Computer Science and Engineering Handbook, pp. 1058–1077. CRC Press.

Franklin, M. J., Carey, M. J., & Livny, M. (1993). Local disk caching for client-server database systems. In Agrawal, R., Baker, S., & Bell, D. A. (Eds.), Proceedings of the VLDB Conference, pp. 641–655.

Gay, J., & Gruenwald, L. (1997). A clustering technique for object oriented databases. In 8th International Conference on Database and Expert Systems Application (DEXA '97), Vol. 1308 of LNCS, pp. 81–90.

Gerlhof, C., Kemper, A., & Moerkotte, G. (1996). On the cost of monitoring and reorganization of object bases for clustering. ACM SIGMOD Record, 25, 28–33.

Gray, J., & Putzolu, G. R. (1987). The 5 minute rule for trading memory for disk accesses and the 10 byte rule for trading memory for cpu time. In In Proceedings of the ACM SIGMOD conference on Management of Data, pp. 395–398.

He, Z., Blackburn, S. M., Kirby, L., & Zigman, J. (2000). Platypus: The design and implementation of a flexible high performance object store. In 9th International Workshop on Persistent Object Systems (POS9), pp. 100–124.

He, Z., & Darmont, J. (2003). DOEF: A dynamic object evaluation framework. In 14th International Conference on Database and Expert Systems Applications (DEXA 03), Prague, Czech Republic, LNCS.

He, Z., Marquez, A., & Blackburn, S. (2000). Opportunistic prioritised clustering framework (OPCF). In International Symposium on Object and Databases, Vol. 1944 of LNCS, pp. 86–100.

Lee, S., Kim, S., & Kim, W. (2000). The bord benchmark for object-relational databases. In 11th International Conference on Database and Expert Systems Applications (DEXA 2000), Vol. 1873 of LNCS, pp. 6–20 London, UK.

Mohan, C., Haderle, D., Lindsay, B., Pirahesh, H., & Schwarz, P. (1992). ARIES: A transaction recovery method supporting fine-granularity locking and partial rollbacks using write-ahead logging. TODS, 17(1), 94–162.

Singhal, V., Kakkad, S. V., & Wilson, P. R. (1992). Texas: An efficient, portable persistent store. In 5th International Workshop on Persistent Object Systems, pp. 11–33.

Tiwary, A., Narasayya, V., & Levy, H. (1995). Evaluation of OO7 as a system and an application benchmark. In OOPSLA '95 Workshop on Object Database Behavior, Benchmarks and Performance.

TPC (2002a). TPC Benchmark C Standard Specification Revision 5.1, Transaction Processing Performance Council.

TPC (2002b). TPC Benchmark W Specification Version 1.8, Transaction Processing Performance Council.

TPC (2003a). TPC Benchmark H Standard Specification Revision 2.1.0, Transaction Processing Council.

TPC (2003b). TPC Benchmark R Standard Specification Revision 2.1.0, Transaction Processing Performance Council.